\documentclass[11pt,a4paper]{article}
\usepackage{geometry}                % See geometry.pdf to learn the layout options. There are lots.
\usepackage{graphicx}
\usepackage{amssymb}
\usepackage{epstopdf}
\usepackage{textcomp}
\usepackage[applemac]{inputenc} %Eingabe deutscher Umlaute
\usepackage{authblk}
\newcommand{\captionfonts}{\small}

\makeatletter  % Allow the use of @ in command names
\long\def\@makecaption#1#2{%
  \vskip\abovecaptionskip
  \sbox\@tempboxa{{\captionfonts #1: #2}}%
  \ifdim \wd\@tempboxa >\hsize
    {\captionfonts #1: #2\par}
  \else
    \hbox to\hsize{\hfil\box\@tempboxa\hfil}%
  \fi
  \vskip\belowcaptionskip}
\makeatother   % Cancel the effect of \makeatletter

\title{Influence of Annealing on the Optical and Scintillation Properties of CaWO$_4$ Single Crystals}

\author[*]{M.~v. Sivers}
\author[*]{C.~Ciemniak}
\author[**]{A.~Erb} 
\author[*]{F.~v. Feilitzsch} 
\author[*]{A.~Gütlein} 
\author[*]{J.-C.~Lanfranchi} 
\author[*]{J.~Lepelmeier} 
\author[*]{A.~Münster}
\author[*]{W.~Potzel} 
\author[*]{S.~Roth} 
\author[*]{R.~Strauss} 
\author[*]{U.~Thalhammer} 
\author[*]{S.~Wawoczny} 
\author[*]{M.~Willers}
\author[*]{A.~Zöller}

\affil[*]{\small{Physik-Department E15, Technische Universität München, James-Franck-Straße 1, 85748 Garching, Germany}}
\affil[**]{Walther-Meißner-Institute for Low Temperature Research, Walther-Meißner-Straße 8, 85748 Garching, Germany}

\date{}                                           % Activate to display a given date or no date

\begin{document}
\maketitle
\begin{abstract}
We investigate the influence of oxygen annealing on the room temperature optical and scintillation properties of CaWO$_4$ single crystals that are being produced for direct Dark Matter search experiments.
The applied annealing procedure reduces the absorption coefficient at the peak position of the scintillation spectrum ($\sim430$\,nm) by a factor of $\sim6$ and leads to an even larger reduction of the scattering coefficient. Furthermore, the annealing has no significant influence on the \emph{intrinsic} light yield.
An additional absorption occurring at $\sim400$\,nm suggests the formation of O$^-$ hole centers.
Light-yield measurements at room temperature where one crystal surface was mechanically roughened showed an increase of the \emph{measured} light yield by $\sim40\,\%$ and an improvement of the energy resolution at 59.5\,keV by $\sim12\,\%$ for the annealed crystal. We ascribe this result to the reduction of the absorption coefficient while the surface roughening is needed to compensate for the also observed reduction of the scattering coefficient after annealing.
\end{abstract}

\section{Introduction}
\label{sec:introduction}
Calcium tungstate (CaWO$_4$) was one of the first known scintillators. Its radioluminescence was discovered by Thomas Edison in 1896 \cite{nature96}. However, it is not widely used in particle physics mainly because of its long luminescence decay times in the order of \textmu s \cite{mikhailik07}. Nevertheless, CaWO$_4$ is an interesting target material for direct Dark Matter search experiments because of its relatively low intrinsic radioactivity and high light yield. In addition, the heavy atomic mass of tungsten ($A=183$) makes it a favorable target for coherent (spin-independent) WIMP-nucleus scattering\footnote{WIMP: Weakly Interacting Massive Particle} because of the expected $A^2$-dependence of the cross section \cite{lewin95}.\\ 
The CRESST-II (Cryogenic Rare Event Search with Superconducting Thermometers) experiment attempts to detect WIMPs using CaWO$_4$ crystals operated as low-temperature detectors \cite{angloher12}. With these detectors an active background discrimination is achieved by the simultaneous detection of phonons and scintillation light produced by particle interactions in the crystal. \\
The most crucial part in the background discrimination ability of a CRESST-II detector module is the amount of detected scintillation light. Therefore, the crystals should have a high intrinsic light yield. The intrinsic light yield is defined as the number of scintillation photons that are produced per MeV of deposited energy. Since the cylindrical crystals have a large size of $\sim40$\,mm diameter and $\sim40$\,mm height also scattering and absorption of scintillation light in the crystal has to be taken into account. This is of particular importance as in a CRESST-II detector module, there is no optical coupling between the scintillating crystal and the light detector. Due to the high symmetry of the cylindrical shape and the high refractive index of CaWO$_4$ (n$\approx1.95$) $\sim50\%$ of all photons will always be internally reflected \cite{carrier89} and are thus trapped inside the crystal. This means that the measured light yield which is defined as the number of detected photons for a given energy is usually much lower than the intrinsic light yield. Simulations have shown that scattering reduces the fraction of trapped light \cite{wahl05} since a random change of the direction prevents a photon from being always internally reflected. In this prospect, the crystals should exhibit a large scattering coefficient and a small absorption coefficient. \\
The EURECA (European Rare Event Calorimeter Array) project is a future ton-scale experiment which combines efforts of cryogenic Dark Matter search in Europe (CRESST, EDELWEISS\footnote{Expérience pour DEtecter Les Wimps En Site Souterrain}, ROSEBUD\footnote{Rare Objects SEarch with Bolometers UndergrounD}) \cite{kraus07}. It is planned that a large part of the target mass will consist of CRESST-like detector modules.\\
For future runs of CRESST and for the EURECA experiment it is crucially important to ensure the availability of CaWO$_4$ crystals, which fulfill the requirements regarding their optical and scintillation properties as well as their radiopurity. It is therefore desirable to have a direct influence on the selection of the raw materials (CaCO$_3$, WO$_3$), the crystal growth and the after-growth treatment. For this reason, a Czochralski furnace dedicated to the growth of CaWO$_4$ single crystals has recently been installed at the Crystal Laboratory of the Technische Universität München (TUM) \cite{ciemniak09}.\\ 
It is known that the thermal annealing under oxygen atmosphere can ameliorate the mechanical, optical and luminescent properties of CaWO$_4$ crystals \cite{shao08,yakovyna04,yakovyna08}. Oxygen annealing is also used to improve the properties of other tungstate scintillators like ZnWO$_4$ and PbWO$_4$ \cite{zhu02, bavykina08}. Especially PbWO$_4$ was extensively studied since it is used for the electromagnetic calorimeter of the CMS (Compact Muon Solenoid) experiment \cite{diemoz07}. High-temperature annealing results in relief of internal stresses. In addition, oxygen annealing can reduce the oxygen deficiency which is present after growth due to the high growth temperature ($\sim1600\,^\circ$C) and the reduced oxygen partial pressure in the growth atmosphere (see Section \ref{sec:sample}).\\
In this work we use a combination of experiments and Monte-Carlo simulations to determine the absorption and scattering coefficients as well as the intrinsic light yield of a CaWO$_4$ crystal at room temperature before and after oxygen annealing. In addition, light-yield measurements at room temperature were conducted to determine the influence of the annealing process on the performance of the crystal as a scintillator. 

\section{Experimental}
\subsection{Sample Preparation}
\label{sec:sample}
For the measurements a cube-shaped crystal of $18\times18\times18$\,mm$^3$ size was used which was produced from an ingot grown by the Czochralski method at the Crystal Laboratory of the TUM. The ingot was grown in a rhodium crucible under constant flow of a mixture of $99\,\%$ argon and $1\,\%$ oxygen. The details of the growth process are described elsewhere \cite{ciemniak09, wmi11}. All surfaces of the crystal were optically polished.
The annealing was carried out at a temperature of $1450\,^\circ$C under constant flow of pure oxygen and lasted 48\,h. The crystal was heated up and cooled down at a rate of $100\,^{\circ}$C/h.  

\subsection{Transmission Measurement}
\label{sec:transmission}
For the transmission measurements a Perkin Elmer LAMBDA 850 UV/VIS spectrophotometer was used. The transmittance was measured for wavelengths from 250-800\,nm in steps of 1\,nm. 
The transmittance $T$ is defined as 
\begin{equation}
T=I_1/I_0
\label{eq:T}
\end{equation} 
where $I_1$ and $I_0$ are the measured intensities with and without the sample in the beam, respectively. 
The attenuation coefficient $\alpha_{att}$ was obtained using the following equation \cite{wahl05}:
\begin{equation}
T=\frac{(1-R)^2\cdot \exp(-\alpha_{att}d)}{1-R^2 \cdot \exp(-2\alpha_{att}d)}
\label{eq:T2}
\end{equation}
\begin{equation}
\Rightarrow\alpha_{att}=-\ln{\left[\frac{-(1-R)^2+\sqrt{(1-R)^4+4T^2R^2}}{2TR^2}\right]}/d
\label{eq:trans}
\end{equation}
Here $d$ is the thickness of the crystal and $R$ its reflectivity.
The denominator of Equation (\ref{eq:T2}) accounts for multiple reflections.\\
CaWO$_4$ (space group: $I4_1/a$) is weakly birefringent ($\delta\approx0.017$ \cite{mindat}), however, in the measurements the crystal was aligned with the beam of light parallel to the optic axis (c-axis) of the crystal so that no birefringence occurred. Therefore, the reflectivity $R$ can be calculated as 
\begin{equation}
R=\frac{(n_o-1)^2}{(n_o+1)^2}
\end{equation}
with $n_o$ being the ordinary refractive index which was calculated by the following dispersion formula \cite{mcgraw94}:  
\begin{equation}
n_o^2-1= \frac{2.5493\cdot (\lambda/\mu m)^2}{(\lambda/\mu m)^2-0.1347^2} + \frac{0.92\cdot(\lambda/\mu m)^2}{(\lambda/\mu m)^2-10.815^2}  
\label{eq:n_cawo}
\end{equation}
where $\lambda$ is the photon wavelength.

\subsection{MCRIM Technique}
\label{sec:MCRIM}
\begin{figure}[htb]
\begin{center}
\includegraphics[scale=0.4]{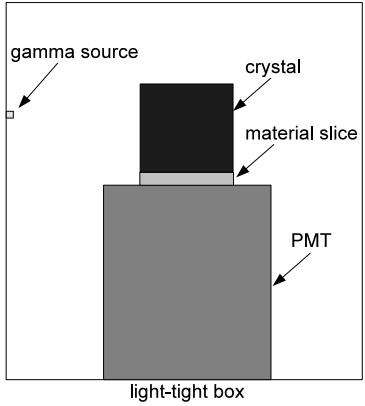}
\caption{Experimental setup of the MCRIM technique. The crystal is placed onto a photomultiplier tube (PMT) and irradiated with a gamma source. A small gap between crystal and PMT offers the possibility to introduce materials with different refractive indices.}
\label{fig:MCRIM}
\end{center}
\end{figure}

The Monte-Carlo Refractive Index Matching (MCRIM) technique is a combination of light-yield measurements and Monte-Carlo (MC) simulations to determine unknown optical and scintillation properties of heavy inorganic scintillators at room temperature \cite{kraus06}. For a crystal with known attenuation coefficient $\alpha_{att}$ it can be used to determine the intrinsic light yield $L_0$ as well as the ratio $B$ of the scattering coefficient $\alpha_{scat}$ and the absorption coefficient $\alpha_{abs}$:
\begin{equation}
B={\alpha_{scat}}/{\alpha_{abs}}
\label{eq:B}
\end{equation}
\begin{equation}
\alpha_{att}=\alpha_{abs}+\alpha_{scat}
\label{eq:a_att}
\end{equation}
Using Equations (\ref{eq:B}) and (\ref{eq:a_att}) one can then calculate the values for the scattering and absorption coefficients.\\
The MCRIM technique is based on the principle that the measured light yield $L_m$ of a scintillating crystal is defined as the product of the intrinsic light yield $L_0$ and the light collection efficiency $\eta$ of the setup in which the crystal is placed. In general, $\eta$ is dependent on a number of parameters such as the geometry of the crystal and the setup, as well as the ratio $B$ and other optical properties of the materials used in the experiment:
\begin{equation}
L_m =\eta L_0
\label{eq:L_m}
\end{equation}
As the measured light yield is dependent on both $B$ and $L_0$, these quantities cannot be determined in a single experiment. However, by taking the ratio $r_{1/2}$ of the measured light yields $L_{m,1}$ and $L_{m,2}$ from two distinctive measurements with different collection efficiencies $\eta_1$ and $\eta_2$ the dependence on $L_0$ can be eliminated:
\begin{equation}
r_{1/2}={L_{m,1}}/{L_{m,2}}={\eta_1}/{\eta_2}
\end{equation}
The experimental setup of the MCRIM technique (see Fig. \ref{fig:MCRIM}) comprises the crystal the properties of which are to be determined and a photomultiplier tube (PMT) placed together in a light-tight box. A small gap between the crystal and the window of the PMT offers the possibility of introducing materials of different refractive indices $n$. The ratio $r_{1/2}$ is obtained by measuring the light yield with two different materials in the gap when the crystal is irradiated with a radioactive source. MC simulations of the setup that iterate the value of $B$ can then be used to match the measured ratio $r_{1/2}$ if all other optical properties are known. In this prospect, the experimental setup is designed to minimize the presence of optical components such as unpolished surfaces and non-ideal reflectors, which are difficult to implement accurately in a simulation. Having determined $B$ through MC matching, the collection efficiency $\eta$ can be calculated by the MC. Using Equation (\ref{eq:L_m}) it is then possible to deduce the intrinsic light yield $L_0$.
A more detailed description of the MCRIM technique is given in Ref. \cite{kraus06}.
The MC simulations of the setup were performed using the GEANT4 (version 4.9.2p01) software \cite{allison06}.
For the measurements an ETL 9305KB photomultiplier was used which was operated in photon counting mode. The PMT pulses were recorded by an Acqiris DC282 digitizer with a sampling rate of 1\,GHz and analyzed off-line. The crystal was irradiated by a $^{241}$Am gamma source (59.5\,keV).
At first a measurement was performed just with air ($n_{air}=1$) between the PMT and the crystal. For the second measurement the gap was filled with an optical gel\footnote{Thorlabs G608N refractive index matching gel} ($n_{gel}=1.46$). The ratio $r_{air/gel}$ is defined as the measured light yield $L_{m,air}$ with air divided by the light yield $L_{m,gel}$ measured with the gap filled with the gel: 
\begin{equation}
r_{air/gel}={L_{m,air}}/{L_{m,gel}}
\end{equation}
%\clearpage{} 

\subsection{Light-Yield Measurements}
\label{sec:L_m}
\begin{figure}[htb]
\begin{center}
\includegraphics[scale=0.4]{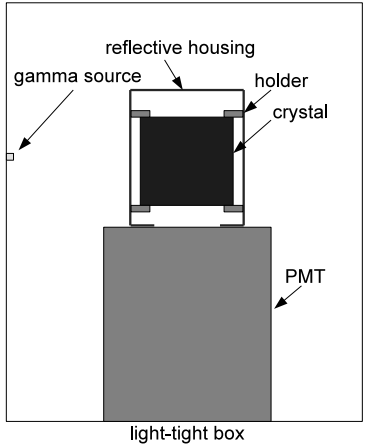}
\caption{Experimental setup of the additional light-yield measurements. The crystal was mounted in a reflective housing and placed onto the PMT window in a geometry which is similar to that of a CRESST-II detector module.}
\label{fig:housing}
\end{center}
\end{figure}

To investigate the influence of the annealing procedure on the performance of the crystal as a scintillator additional light-yield measurements were performed at room temperature.
In these measurements the crystal was mounted in an aluminum housing (see Fig. \ref{fig:housing}) which was then placed onto the PMT of the setup described in Section \ref{sec:MCRIM}. The housing is covered with a highly reflective foil\footnote{3M Radiant Mirror Film VM2000} on the inside and the crystal is supported by small holders made of polyethylene. In this configuration there is a small gap between the crystal and the side walls of the reflective housing as well as between the crystal and the PMT window (see Fig. \ref{fig:housing}). In this way the geometry is similar to that of a CRESST-II detector module \cite{angloher12}.
It is known that surface roughening can increase the light collection especially when there is no optical coupling between crystal and light detector \cite{strul02}. This increased light collection can be ascribed to the fact that each photon will hit the roughened surface at a random incident angle which prevents it from being always internally reflected. Therefore surface roughening decreases the fraction of light trapped in the crystal similar to scattering. Furthermore, a rough surface can show strong distributions of transmitted light for angles of incidence larger than the critical angle where no transmitted light would be observed at a specular surface \cite{nieto-vesperinas92}. In CRESST-II, a significant improvement of the energy resolution was obtained by mechanical roughening of the crystal surface that is facing the light detector using silicon carbide powder with a grain size of $\sim9$\,\textmu m \cite{ninkovic05}.
Therefore, in the measurements that are described in this section the crystal surface that is facing the PMT was roughened by applying the identical procedure.\footnote{The roughening procedure was also reapplied after annealing.}
The crystal was again irradiated with a $^{241}$Am gamma source (59.5\,keV). 
%\clearpage{}

\section{Results and Discussion}
\subsection{Transmission Measurement}
\label{sec:transmission_results}
\begin{figure}[htb]
\begin{center}
\includegraphics[scale=0.6]{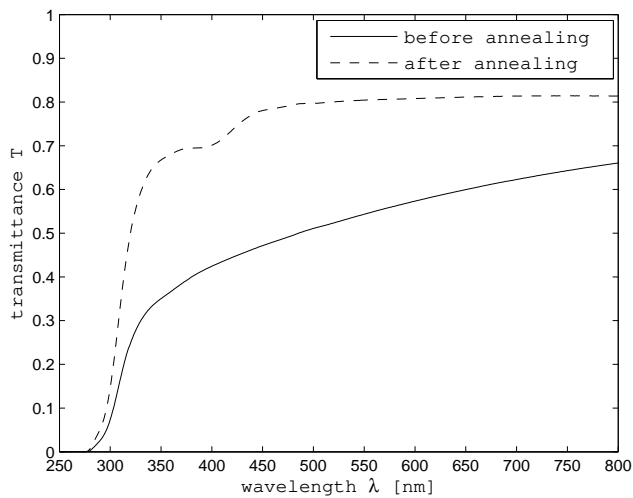}
\caption{Transmittance $T$ of the CaWO$_4$ crystal before and after annealing versus photon wavelength $\lambda$. The pronounced absorption around 400\,nm in the annealed crystal is probably caused by O$^-$ hole centers.}
\label{fig:wl_trans}
\end{center}
\end{figure}

\begin{figure}[htb]
\begin{center}
\includegraphics[scale=0.6]{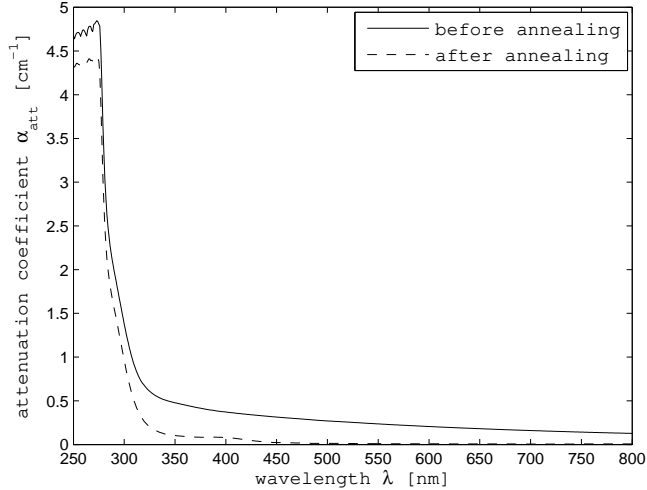}
\caption{Attenuation coefficient $\alpha_{att}$ of the CaWO$_4$ crystal before and after annealing versus photon wavelength $\lambda$.}
\label{fig:wl_a_att}
\end{center}
\end{figure}

Fig. \ref{fig:wl_trans} shows the measured transmittance $T$ (see Equation \ref{eq:T}) of the crystal before and after annealing in dependence of the photon wavelength $\lambda$. 
It can be seen that the transmittance is considerably increased after the annealing process. Fig. \ref{fig:wl_a_att} shows the attenuation coefficient $\alpha_{att}$ calculated from the measured transmittance (see Equation \ref{eq:trans}) versus the photon wavelength. Tab. \ref{tab:a_att} summarizes the values of the attenuation coefficient at a photon wavelength of 430\,nm which corresponds to the approximate peak position of the CaWO$_4$ scintillation spectrum at room temperature \cite{mikhailik04}.\\
The error for the attenuation coefficient is given at $95\,\%$ CL and was calculated from the data of four independent transmission measurements. However, it has to be pointed out that, in addition, there are some systematic uncertainties in the value of $\alpha_{att}$ because the refractive index of the crystal was not directly measured. Furthermore, surface irregularities can cause errors in the measured transmittance in the order of $\sim20\,\%$ \cite{wahl05}.
\begin{table}[htb]
\begin{center}
\begin{tabular}{|c|c|}
\hline
& $\alpha_{att}$ [cm$^{-1}$] at 430\,nm \\
\hline
before annealing & $0.335\pm0.026$  \\
after annealing & $0.039\pm0.010$ \\
\hline
\end{tabular}
\caption{Value of the attenuation coefficient $\alpha_{att}$ at 430\,nm of the CaWO$_4$ crystal before and after annealing. The wavelength corresponds to the approximate peak position of the scintillation spectrum at room temperature. The errors are given at $95\,\%$ CL and were determined from the data of four independent transmission measurements.}
\label{tab:a_att}
\end{center}
\end{table}
Tab. \ref{tab:a_att} shows that the attenuation coefficient at 430\,nm is decreased by a factor of $\sim8$ after the annealing procedure. This can be attributed to the removal of crystal defects which otherwise cause scattering or absorption of the photons. 
Scattering centers in CaWO$_4$ crystals include solid and gaseous inclusions \cite{nassau62, cockayne64}. 
Solid inclusions can be caused by the incorporation of rhodium from the crucible \cite{cockayne64a, shappirio63} while gaseous inclusions probably result from oxygen separation from WO$_4^{2-}$ complexes \cite{oeder76}.
It is, however, unclear if inclusions that are present in the crystal can be removed by annealing.
Dislocations that are present in the crystal can act as dielectric inhomogeneities and hence also scatter light \cite{kesavamoorthy83}. The formation of dislocations can result from temperature gradients after growth in the cooling crystal \cite{cockayne64}.
During annealing dislocations become mobile and can be removed from the crystal or form low-angle boundaries \cite{levinstein63, cockayne64}. This may be a reason for the observed reduction of the attenuation coefficient. 
In addition, the main intrinsic defect type in CaWO$_4$ crystals are oxygen vacancies \cite{shao08} and their formation is facilitated by the reduced oxygen partial pressure in the growth atmosphere (see Section \ref{sec:sample}). These oxygen vacancies can be filled during annealing under oxygen atmosphere which can also reduce optical scattering and absorption. \\
In Fig. \ref{fig:wl_trans} one can observe a pronounced absorption around 400\,nm in the annealed crystal. According to Ref. \cite{yakovyna04} oxygen can also be incorporated on interstitial sites during annealing which leads to absorption bands at $\sim400$\,nm and $\sim310$\,nm.
However, it is also known that for PbWO$_4$ an absorption band at $\sim420\,$nm can be induced after oxygen annealing \cite{nikl96, nikl97}. This absorption band is commonly ascribed to O$^-$ hole centers \cite{nikl96}. The filling of oxygen vacancies during annealing might lead to the creation of O$^-$ centers to maintain local charge balance. It is known that for CaWO$_4$ the localization of holes at one oxygen of the WO$_4$ tetrahedron is the dominant trapping mechanism \cite{nikl08}. In this way the O$^-$ hole center may be a more natural explanation for the observed absorption band.\\
The vanishing transmittance below $\sim300$\,nm in both curves of Fig. \ref{fig:wl_trans} is due to intrinsic absorption corresponding to the excitation of electrons from the valence band to the conduction band of CaWO$_4$ \cite{mikhailik04}.  
%\clearpage{}

\subsection{MCRIM Technique}
\label{sec:MCRIM_results}
Tab. \ref{tab:MCRIM_Lm} shows the results of the light-yield measurements from the MCRIM technique before and after annealing of the crystal. 
The results for the intrinsic light yield $L_0$, the absorption coefficient $\alpha_{abs}$ and the scattering coefficient $\alpha_{scat}$, that were determined from the matching of the values of Tab. \ref{tab:MCRIM_Lm} to MC simulations, are summarized in Tab. \ref{tab:MCRIM}.
All errors are given at $95\,\%$ CL and were determined from the data of four independent light-yield measurements and the uncertainty of the attenuation coefficient (see Tab. \ref{tab:a_att}).
\begin{table}[htb]
\begin{center}
\begin{tabular}{|c|c|c|}
\hline
 & $L_{m,air}$ [p.e.] at 59.5\,keV & $L_{m,gel}$ [p.e.] at 59.5\,keV\\
\hline
before annealing & $39.9\pm0.7$ & $79.3\pm1.7$  \\
after annealing & $39.2\pm0.4$  & $92.0\pm1.0$ \\
\hline
\end{tabular}
\caption{Results of the light-yield measurements from the MCRIM technique performed with a $^{241}$Am gamma source (59.5\,keV). $L_{m,air}$ and $L_{m,gel}$ denote the measured light yields in photoelectrons (p.e.) with the gap between crystal and PMT (see Fig. \ref{fig:MCRIM}) filled with air and gel, respectively. All errors are given at $95\,\%$ CL and were determined from the data of four independent light-yield measurements.}
\label{tab:MCRIM_Lm}
\end{center}
\end{table}
\begin{table}[htb]
\begin{center}
\begin{tabular}{|c|c|c|c|}
\hline
 & $L_0$ [ph/MeV] at 59.5\,keV &$\alpha_{abs}$ [cm$^{-1}$] &$\alpha_{scat}$ [cm$^{-1}$]  \\
\hline
before annealing & $24800\pm3300$ & $0.231\pm0.051$ & $0.104\pm0.048$ \\
after annealing & $20600\pm900$ & $0.036\pm0.010$ & $0.004\pm0.003$ \\
\hline
\end{tabular}
\caption{Results of the MCRIM technique for the intrinsic light yield $L_0$ at 59.5\,keV, the absorption coefficient $\alpha_{abs}$ and the scattering coefficient $\alpha_{scat}$. All errors are given at $95\,\%$ CL and were determined from the data of four independent light-yield measurements and the uncertainty of the attenuation coefficient (see Tab. \ref{tab:a_att}).}
\label{tab:MCRIM}
\end{center}
\end{table}
The values show that the annealing procedure leads to a decrease of the absorption coefficient of the crystal by a factor of $\sim6$ and an even larger reduction of the scattering coefficient. The measurements also suggest that the attenuation of scintillation light in the annealed crystal is dominated by absorption. However, as the errors of the scattering and absorption coefficients are rather large no definitive statements can be made.\\
As mentioned in Section \ref{sec:introduction} scattering can increase the light collection due to the reduction of light being trapped inside the crystal. This explains the fact that the measured light yield $L_{m,air}$ with no optical coupling between crystal and PMT is not increased after annealing despite the reduction of the absorption coefficient. In contrast to this, the measured light yield $L_{m,gel}$ with optical coupling - where trapping is far less prominent - increases, as it is hardly influenced by the reduction of the scattering coefficient. Since in a CRESST-II detector module there is also no optical coupling to the light detector (see Section \ref{sec:L_m}) it would be desirable to increase the transmittance of the crystal without removing the scattering centers. 
However, the reduction of trapped light can also be accomplished by surface roughening (see Sections \ref{sec:L_m} and \ref{sec:L_m_results}).\\
The value of the intrinsic light yield $L_0$ shows a small, although not significant, decrease after the annealing procedure. 
This may result from the formation of defects which act as quenching centers, i.e. they are responsible for energy absorption followed by non-radiative decay.\\
We note that the values of the intrinsic light yield from Tab. \ref{tab:MCRIM} are comparable to the value of $22700\pm1100$\,ph/MeV which was determined in Ref. \cite{kraus06} for a CaWO$_4$ crystal produced by the Institute of Materials SRC "Carat" (Ukraine).\footnote{The "Carat" institute is one of the current suppliers for CRESST-II detector crystals.}
%\clearpage{} 

\subsection{Light-Yield Measurements}
\label{sec:L_m_results}
\begin{figure}[htb]
\begin{center}
\includegraphics[scale=0.6]{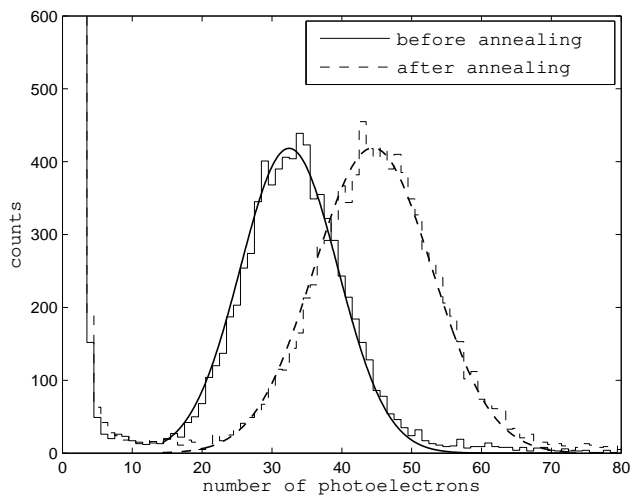}
\caption{Spectra of $^{241}$Am gamma rays (59.5\,keV) measured with the roughened CaWO$_4$ crystal before and after annealing. The measured light yield (peak position) and energy resolution (FWHM/peak position) were determined from a Gaussian fit to the photopeak.}
\label{fig:L_m}
\end{center}
\end{figure}

Fig. \ref{fig:L_m} shows two $^{241}$Am spectra measured with the CaWO$_4$ crystal before and after annealing. As mentioned in Section \ref{sec:L_m} the crystal surface that was facing the PMT had been mechanically roughened for these measurements as it is done for CRESST-II detectors.\\
In Tab. \ref{tab:L_m} we show the mean values and errors ($95\,\%$ CL) for the measured light yield $L_m$ as determined from the peak position, and for the energy resolution as derived from the ratio of the peak width (FWHM) and peak position. The results were calculated from the data of four independent light-yield measurements.
\begin{table}[htb]
\begin{center}
\begin{tabular}{|c|c|c|}
\hline
 & $L_m$ [p.e.] at 59.5\,keV & resolution [\%] at 59.5\,keV\\
\hline
before annealing & $31.0\pm0.6$ & $52.2\pm0.6$  \\
after annealing & $43.3\pm0.5$  & $46.0\pm1.0$ \\
\hline
 \end{tabular}
\caption{Measured light yield $L_m$ in photoelectrons (p.e.) and energy resolution at 59.5\,keV of the roughened crystal before and after annealing. The mean values and errors ($95\,\%$ CL) were determined from the data of four independent light-yield measurements.}
\label{tab:L_m}
\end{center}
\end{table}
In contrast to the measured light yield $L_{m,air}$ of the polished crystal (see Tab. \ref{tab:MCRIM_Lm}), the measured light yield $L_m$ of the roughened crystal (see Tab. \ref{tab:L_m}) is increased by $\sim40\,\%$ and the energy resolution is improved by $\sim12\,\%$ after annealing despite of a large reduction of the scattering coefficient (see Tab. \ref{tab:MCRIM}).\footnote{The fact that the measured light yield $L_m$ of the roughened crystal before annealing is lower than the value $L_{m,air}$ of the polished crystal (see Tab. \ref{tab:MCRIM_Lm}) results from the different geometries of both setups (see Figs. \ref{fig:MCRIM} and \ref{fig:housing}). In particular, the additional holders of the crystal and the smaller surface area where photons can escape the reflective housing and be detected lead to a lower light collection.}
As mentioned in Section \ref{sec:L_m} surface roughening can reduce the fraction of scintillation light that is trapped in the crystal similar to scattering. We thus infer that the effects of the roughened surface are dominant over those caused by scattering in the crystal. In this way, the reduction of the scattering coefficient can be compensated by the roughened surface and the observed improvement of the measured light yield and energy resolution after annealing can be explained by the reduction of the absorption coefficient (see Tab. \ref{tab:MCRIM}).
To support this argument we have used the MC simulations mentioned in Section \ref{sec:MCRIM} to determine the fraction of trapped light in the crystal before and after annealing. This was done for the cases that all surfaces of the crystal are polished and that one of the surfaces is roughened. In the simulation a rough surface is a collection of micro-facets with a Gaussian distribution of facet slopes \cite{levin96}. The standard deviation of this distribution that was implemented in the simulation was determined from the measured surface profile of a roughened crystal.
\begin{table}[htb]
\begin{center}
\begin{tabular}{|c|c|c|}
\hline
 & all surfaces polished & one surface roughened \\
\hline
before annealing  & $62.8\pm5.2\%$ & $59.8\pm4.3\%$  \\
after annealing & $56.8\pm3.2\%$  & $35.9\pm4.9\%$ \\
\hline
 \end{tabular}
\caption{Simulation of the fraction of light that is trapped inside the crystal before and after annealing. The simulation was conducted with either all surfaces polished or with one surface roughened. The mean values and errors (95\% CL) were determined from the uncertainties on the absorption and scattering coefficients.}
\label{tab:trapping}
\end{center}
\end{table}
The results of the simulations are shown in Tab. \ref{tab:trapping}.
It can be seen that the polished crystal shows no significant change of the fraction of trapped light after annealing. However, with one surface roughened the fraction of trapped light is reduced by $\sim40\%$ after annealing. This agrees well with the measured increase of $L_m$ by $\sim40\%$ after annealing of the roughened crystal (see Tab. \ref{tab:L_m}). Therefore the simulations support the interpretation that the effects of surface roughening are dominant and compensate the decrease of the scattering coefficient.\\
We note that a similar result was found for ZnWO$_4$ crystals which have shown an improvement of the measured light yield in a similar setup by $\sim30\%$ after oxygen annealing \cite{bavykina08}.
%\clearpage{}

\section{Summary and Conclusion}
We have investigated the room temperature optical and scintillation properties of an $18\times18\times18$\,mm$^3$ CaWO$_4$ crystal that was produced by the Czochralski method at the Crystal Laboratory of the TUM in a setup which is dedicated to the growth of CaWO$_4$ crystals for the direct Dark Matter search experiments CRESST and EURECA. Using transmission measurements and the MCRIM technique we determined the absorption and scattering coefficients as well as the intrinsic light yield of the crystal. The measurements were performed before and after the crystal was annealed under a constant flow of pure oxygen at 1450\,$^\circ$C for 48\,h.\\ 
It was shown that annealing decreases the absorption coefficient of the crystal by a factor of $\sim6$ and leads to an even larger reduction of the scattering coefficient. This result is likely to be explained by the removal of dislocations and oxygen vacancies that are present in the crystal directly after growth. Furthermore, the applied annealing procedure has no significant influence on the intrinsic light yield. 
An additional absorption occurring around 400\,nm in the transmission curve suggests the formation of O$^-$ hole centers during annealing.
Further light-yield measurements at room temperature were performed where one crystal surface was mechanically roughened as it is done for CRESST-II detectors. The results showed an increase of the measured light yield by $\sim40\,\%$ and an improvement of the energy resolution at 59.5\,keV by $\sim12\,\%$ for the annealed crystal. We ascribe this result to the reduction of the absorption coefficient while the surface roughening is needed to compensate for the also observed reduction of the scattering coefficient after annealing.\\
Our measurements have shown that the oxygen annealing of CaWO$_4$ crystals can lead to a significant improvement of their performance as a scintillator at room temperature. This behaviour has also to be confirmed at low temperatures ($\sim$10mK) in the operating range of CRESST/EURECA detector modules.
\section*{Acknowledgements}
We thank D. Wahl and V. Mikhailik for helpful discussions on the MCRIM technique.
This work has been supported by funds of the Deutsche Forschungsgemeinschaft DFG (Transregio 27: Neutrinos and Beyond), the Excellence Cluster (Origin and Structure of the Universe) and the Maier-Leibnitz-La\-bo\-ra\-to\-ri\-um (Garching). 


\bibliographystyle{prsty}
\bibliography{BibtexDatabase}

\end{document}